\documentclass[runningheads]{llncs}
\usepackage{multirow}
\usepackage{bbding} 
\usepackage[misc]{ifsym}
\usepackage{verbatim}
\usepackage{tabularx}
\usepackage[colorlinks,linkcolor=green, anchorcolor=blue, citecolor=blue, urlcolor=blue]{hyperref}
\usepackage{graphicx}
\usepackage{amsmath}
\usepackage{siunitx}
\usepackage{bm}

\begin{document}
\title{H$^2$NF-Net for Brain Tumor Segmentation using Multimodal MR Imaging: 2nd Place Solution to BraTS Challenge 2020 Segmentation Task}
 
\titlerunning{H$^2$NF-Net for Brain Tumor Segmentation} 

\author{Haozhe Jia\inst{1,2,4} \and Weidong Cai \inst{3} \and Heng Huang \inst{4,5} \and Yong Xia\textsuperscript{1,2(\Letter)}}

\authorrunning{Jia et al.}

\institute{\textsuperscript{1} Research \& Development Institute of Northwestern Polytechnical University in Shenzhen, Shenzhen 518057, China\\ \email{yxia@nwpu.edu.cn}\\ 
\textsuperscript{2} National Engineering Laboratory for Integrated Aero-Space-Ground-Ocean Big Data Application Technology, School of Computer Science and Engineering, Northwestern Polytechnical University, Xi'an 710072, China\\
\textsuperscript{3} School of Computer Science, University of Sydney, Sydney, NSW 2006, Australia\\
\textsuperscript{4} Department of Electrical and Computer Engineering, University of Pittsburgh, Pittsburgh, PA 15261, USA\\
\textsuperscript{5} JD Finance America Corporation, California, CA 94043, USA}
\maketitle

\begin{abstract}
In this paper, we propose a Hybrid High-resolution and Non-local Feature Network (H$^2$NF-Net) to segment brain tumor in multimodal MR images. 
Our H$^2$NF-Net uses the single and cascaded HNF-Nets to segment different brain tumor sub-regions and combines the predictions together as the final segmentation.
We trained and evaluated our model on the Multimodal Brain Tumor Segmentation Challenge (BraTS) 2020 dataset. 
The results on the test set show that the combination of the single and cascaded models achieved average Dice scores of 0.78751, 0.91290, and 0.85461, as well as Hausdorff distances ($95\%$) of 26.57525, 4.18426, and 4.97162 for the enhancing tumor, whole tumor, and tumor core, respectively. 
Our method won the second place in the BraTS 2020 challenge segmentation task out of nearly 80 participants.

\end{abstract}

\section{Introduction}
Brain gliomas are the most common primary brain malignancies, which generally contain heterogeneous histological sub-regions, i.e. edema/invasion, active tumor structures, cystic/necrotic components, and non-enhancing gross abnormality. 
Accurate and automated segmentation of these intrinsic sub-regions using multimodal magnetic resonance (MR) imaging is critical for the potential diagnosis and treatment of this disease. 
To this end, the multimodal brain tumor segmentation challenge (BraTS) has been held for many years, which provides a platform to evaluate the state-of-the-art methods for the segmentation of brain tumor sub-regions \cite{bakas2017segmentation1,bakas2017segmentation2,bakas2017advancing,bakas2018identifying,menze2014multimodal}.

With deep learning being widely applied to medical image analysis, fully convolutional network (FCN) based methods have been designed for this segmentation task and have shown convincing performance in previous challenges. 
Kamnitsas \textit{et al}. \cite{kamnitsas2017efficient} constructed a 3D dual pathway CNN, namely DeepMedic, which simultaneously processes the input image at multiple scales with a dual pathway architecture so as to exploit both local and global contextual information. 
DeepMedic also uses a 3D fully connected conditional random field to remove false positives.
In \cite{isensee2018no}, Isensee \textit{et al}. achieved outstanding segmentation performance using a 3D U-Net with instance normalization and leaky ReLU activation, in conjunction with a combination loss function and a region-based training strategy. 
In \cite{myronenko20183d}, Myronenko \textit{et al}. incorporated a variational auto-encoder (VAE) based reconstruction decoder into a 3D U-Net to regularize the shared encoder, and achieved the 1st place segmentation performance in BraTS 2018. 
In BraTS 2019, Jiang \textit{et al}. \cite{jiang2019two} proposed a two-stage cascaded U-Net to segment the brain tumor sub-regions from coarse to fine, where the second-stage model has more channel numbers and uses two decoders so as to boost performance. 
This method achieved the best performance in the BraTS 2019 segmentation task.
In our previous work \cite{jia2020learning}, we proposed a High-resolution and Non-local Feature Network (HNF-Net) to segment brain tumor in multimodal MR images. 
The HNF-Net is constructed based mainly on the parallel multi-scale fusion (PMF) module, which can maintain strong high-resolution feature representation and aggregate multi-scale contextual information.
The expectation-maximization attention (EMA) module is also introduced to the model to enhance the long-range dependent spatial contextual information at the cost of acceptable computational complexity.

In this paper, we further propose a Hybrid High-resolution and Non-local Feature Network (H$^2$NF-Net) for this challenging task. 
Compared to original HNF-Net, the proposed H$^2$NF-Net adds a two-stage cascaded HNF-Net and uses the single and cascaded models to segment different brain tumor sub-regions.
We evaluated the proposed method on the BraTS 2020 challenge dataset. 
In addition, we also introduced the detailed implementation information of our second place solution to BraTS 2020 challenge segmentation task.

\begin{figure}[tb]
\centering
\includegraphics[width=1\textwidth]{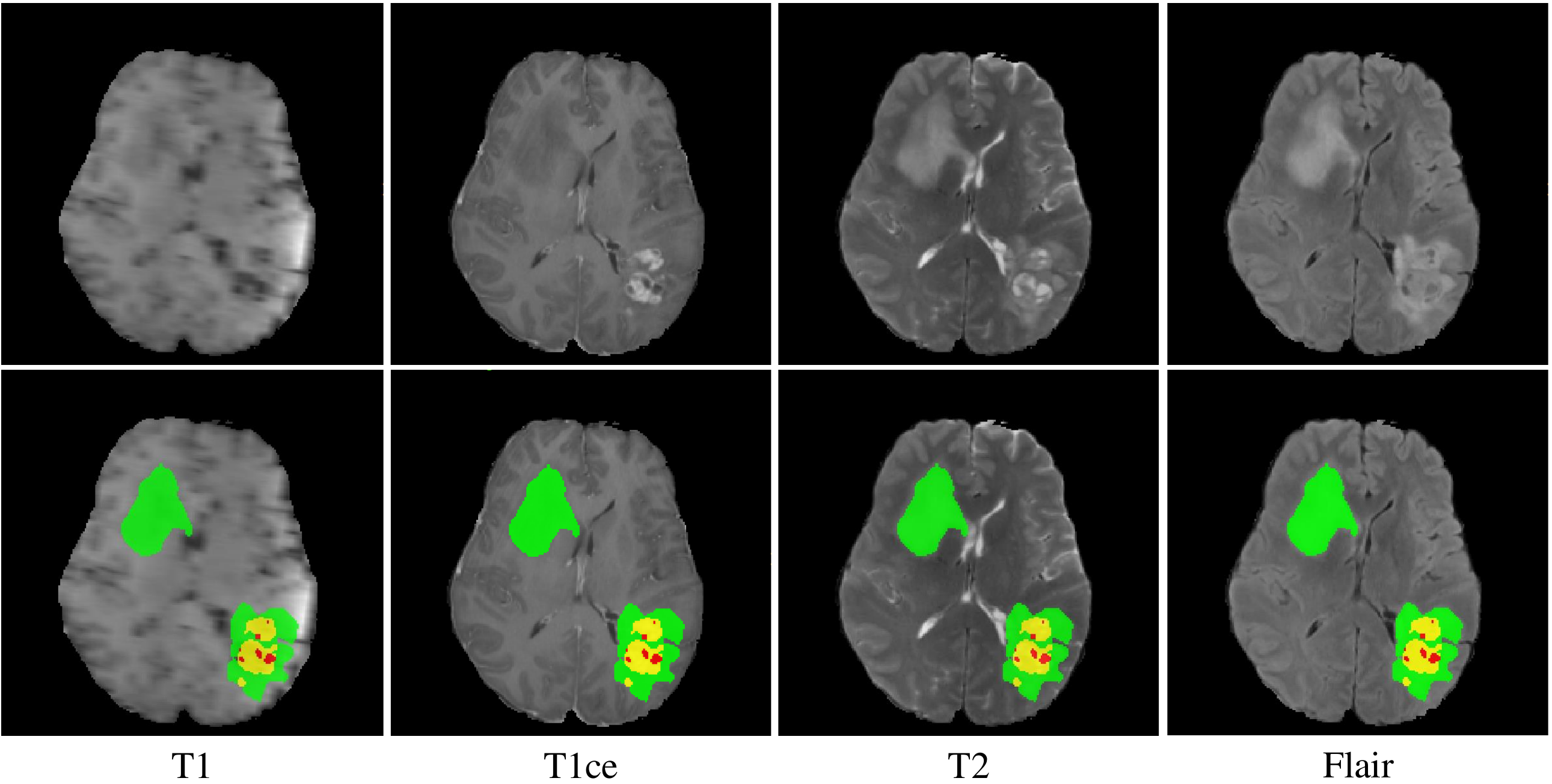}
\caption{Example scans and corresponding annotations of different modalities. The NCR/NET, ED, and ET regions are highlighted in red, green, and yellow, respectively.}
\label{fig:fig1}
\end{figure}

\section{Dataset}
The BraTS 2020 challenge dataset \cite{bakas2017segmentation1,bakas2017segmentation2,bakas2017advancing,bakas2018identifying,menze2014multimodal} contains 369 training, 125 validation and 166 test multimodal brain MR studies. 
Each study has four MR images, including T1-weighted (T1), post-contrast T1-weighted (T1ce), T2-weighted (T2), and fluid attenuated inversion recovery (Flair) sequences, as shown in Fig. \ref{fig:fig1}. 
All MR images have the same size of $240\times240\times155$ and the same voxel spacing of $1\times1\times1mm^3$. 
For each study, the enhancing tumor (ET), peritumoral edema (ED), and necrotic and non-enhancing tumor core (NCR/NET) were annotated on a voxel-by-voxel basis by experts. 
The annotations for training studies are publicly available, and the annotations for validation and test studies are withheld for online evaluation and final segmentation competition, respectively.

\section{Method}
In this section, we introduce the structure of our H$^2$NF-Net. Despite the details have been introduced in our previous work \cite{jia2020learning}, to make the proposed H$^2$NF-Net self-consistent, here we still give a brief introduction of HNF-Net and its two key modules. 
Also we provide the details of the two-stage cascaded HNF-Net.
We now delve into the details of each part.

\subsection{Single HNF-Net}
\begin{figure}[t]
\centering
\includegraphics[width=1\textwidth]{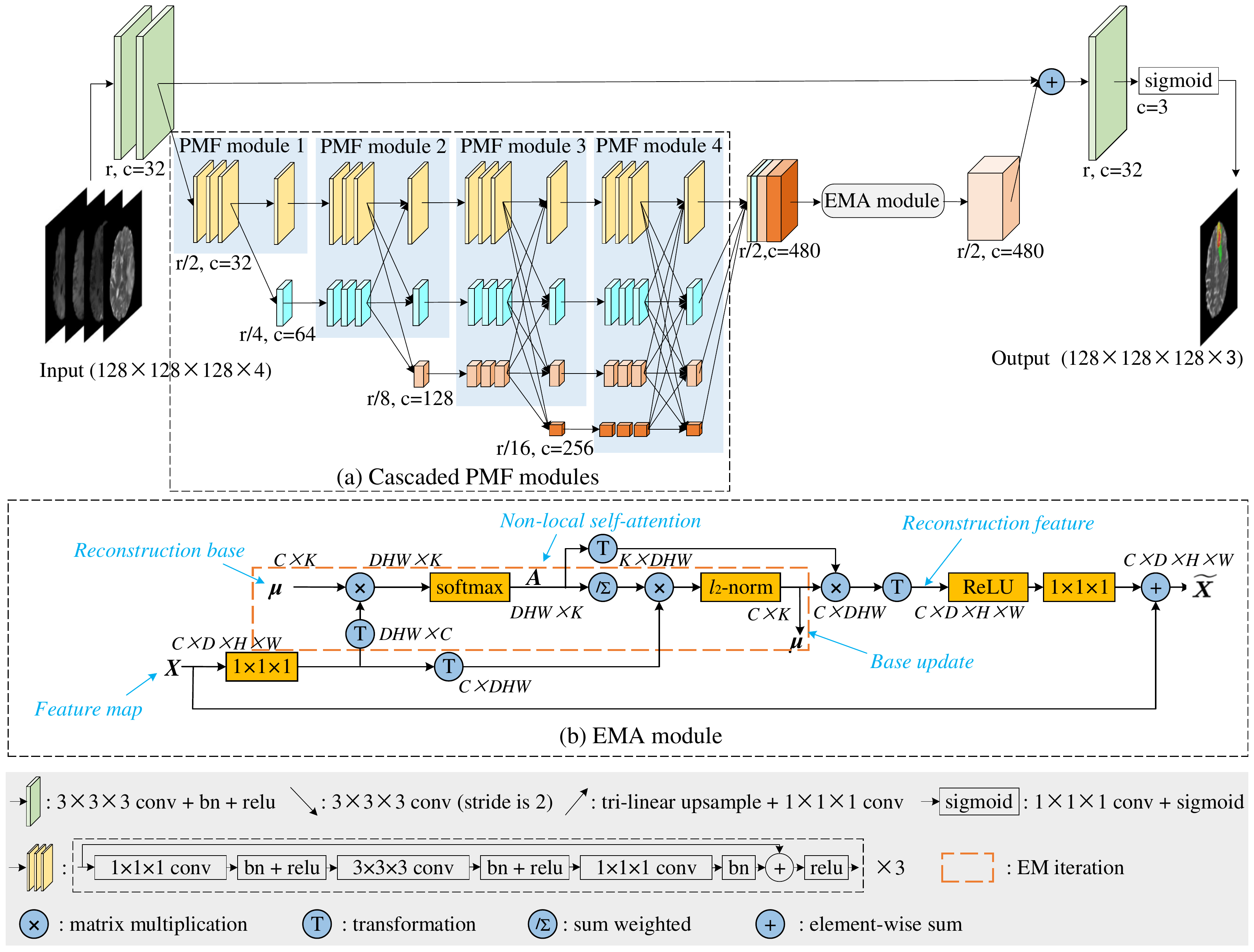}
\caption{Architecture of the single HNF-Net, PMF module and EMA module. r denotes the original resolution and c denotes the channel number of feature maps. All downsample operations are achieved with 2-stride convolutions, and all upsample operations are achieved with joint $1\times 1\times 1$ convolutions and tri-linear interpolation. It is noted that, since it is inconvenient to show 4D feature maps ($C\times D\times H \times W$) in the figure, we show all feature maps without depth information, and the thickness of each feature map reveals its channel number.}
\label{fig:fig2}
\end{figure}

The single HNF-Net has an encoder-decoder structure. For each study, four multimodal brain MR sequences are first concatenated to form a four-channel input and then processed at five scales i.e., $r$, $1/2r$, ... 1/16$r$, highlighted in green, yellow, blue, pink, and orange in the upper figure of Fig. \ref{fig:fig2}, respectively.
At the original scale $r$, there are four convolutional blocks, two for encoding and the other two for decoding. The connection from the encoder to the decoder skips the processing at other scales so as to maintain the high resolution and spatial information in a long-range residual fashion. At other four scales, four PMF modules are jointly used as a high-resolution and multi-scale aggregated feature extractor.
At the end of the last PMF module, the output feature maps at four scales are first recovered to the 1/2$r$ scale and then concatenated as mixed features. Next, the EMA module is used to efficiently capture long-range dependent contextual information and reduce the redundancy of the obtained mixed features. Finally, the output of the EMA module is recovered to original scale $r$ and 32 channels via $1\times 1\times 1$ convolutions and upsampling and then added to the full-resolution feature map produced by the encoder for the dense prediction of voxel labels. All downsample operations are achieved with 2-stride convolutions, and all upsample operations are achieved with joint $1\times 1\times 1$ convolutions and tri-linear interpolation.

\begin{figure}[t]
\centering
\includegraphics[width=1\textwidth]{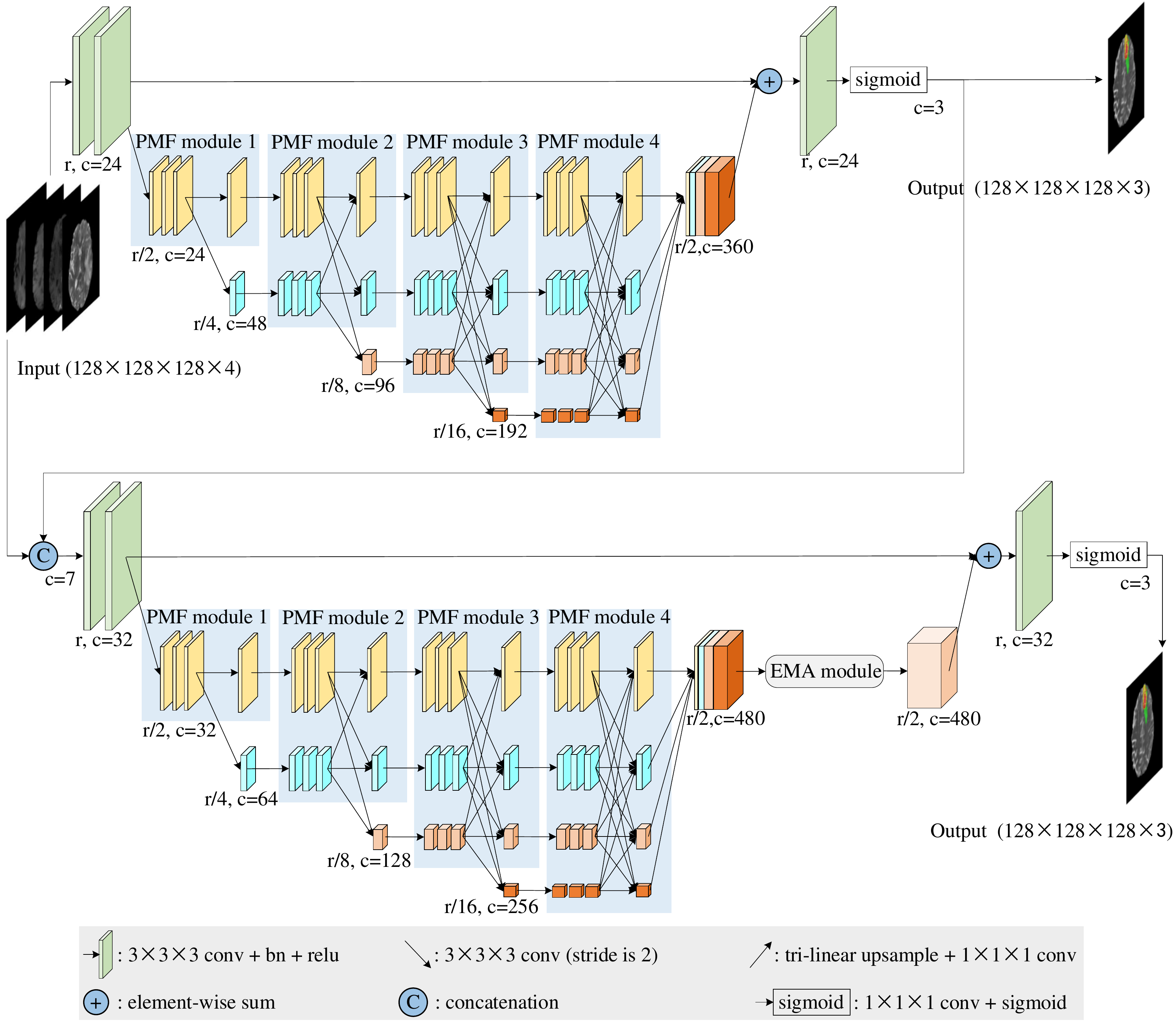}
\caption{Architecture of the cascaded HNF-Net. r denotes the original resolution and c denotes the channel number of feature maps.}
\label{fig:fig3}
\end{figure}

\noindent\textbf{PMF module.}
Given that learning strong high-resolution representation is essential for small object segmentation, high-resolution features are maintained throughout the segmentation process in recent solutions \cite{pohlen2017full,saxena2016convolutional,sun2019high}, and it has been shown that this strategy contributes to convincing performance. Based on this, the PMF module is designed to have two parts, the parallel multi-scale convolutional block and fully connected fusion block. The former has a set of parallel branches similar to the group convolution, and each branch is built with repeated residual convolutional blocks at a specific scale. The latter fuses all output features of the parallel multi-scale convolutional block in a parallel but fully connected fashion, where each branch is the summation of the output features of all resolution branches. Thus, in each PMF module, the parallel multi-scale convolutional block can fully exploit multi-resolution features but maintain high-resolution feature representation, and the fully connected fusion block can aggregate rich multi-scale contextual information. Moreover, we cascade multiple PMF modules, in which the number of branches increases progressively with depth, as shown in Fig. \ref{fig:fig2}(a). As a result, from the perspective of the highest resolution stage, its high-resolution feature representation is boosted with repeated fusion of multi-scale low-resolution representations. Meanwhile, the cascade PMF modules can be regarded as an ensemble of several U-shape sub-networks with different depths and widths, which tends to further reduce the semantic gap of the features at different depths.

\noindent\textbf{EMA module.}
It has been proved that Non-local self-attention mechanism can help to aggregate contextual information from all spatial positions and capture long-range dependencies \cite{fu2019dual,wang2018non,zhao2018psanet}. However, the potential high computational complexity makes it hard to be applied to 3D medical image segmentation tasks. Hence, we introduce the EMA module \cite{li2019expectation} to our tumor segmentation model, aiming to incorporate a lightweight Non-local attention mechanism into our model. The main concept of the EMA module is operating the Non-local attention on a set of feature reconstruction bases rather than directly achieving this on the high-resolution feature maps. Since the reconstruction bases have much less elements than the original feature maps, the computation cost of the Non-local attention can be significantly reduced. As shown in the lower figure of Fig. \ref{fig:fig2} (b), the shape of the input feature maps \textit{X} are $C\times D\times H\times W$, where $C$ is the channel number. The dotted box denotes the EM algorithm operation, where the Non-local spatial attention maps \textit{A} and the bases \textit{$\mu$} are generated alternately as the E step and M step, respectively. After convergence, we can use the obtained \textit{A} and \textit{$\mu$} to generate reconstructed feature maps $\widetilde{X}$. At both the beginning and ending of the EM algorithm operation, the $1\times1\times1$ convolutions are adopted to change the channel number. In addition, to avoid overfitting, we further sum $\widetilde{X}$ with $X$ in a residual fashion.

\subsection{Cascaded HNF-Net}
Inspired by the previous work \cite{jiang2019two,wu2018multiscale}, we further construct a two-stage cascaded version of our HNF-Net, with the structure shown in Fig. \ref{fig:fig3}. Compared to the single HNF-Net, the first stage network of the cascaded HNF-Net is narrower and has no EMA module. The second network has the same structure with the single HNF-Net, but receives the concatenation of the original image block and the prediction of the first network as the input. The two-stage cascaded HNF-Net is trained in an end-to-end fashion, and the deep supervision is also added for the output of the first stage network for a stable training. Given that (1) the segmentation performance of three sub-regions, namely ET, tumor core (TC, the union of ET and NCR/NET), and whole tumor (WT, the union of all sub-regions) is evaluated separately in BraTS 2020 and (2) the single HNF-Net shows better segmentation performance on ET and WT, while the cascaded HNF-Net has better results on TC on the validation set, we use the single HNF-Net to segment ED and ET and use the cascaded HNF-Net to segment NCR/NET in the practical use of our H$^2$NF-Net. 

\section{Experiments and Results}
\subsection{Implementation Details}
\noindent\textbf{Pre-processing.}
Following our previous work \cite{jia2020learning}, we performed a set of pre-processing on each brain MR sequence independently, including brain stripping, clipping all brain voxel intensity with a window of [0.5\%-99.5\%], and normalizing them into zero mean and unit variance. 

\noindent\textbf{Training.}
In the training phase, we randomly cropped the input image into a fixed size of $128\times128\times128$ and concatenated four MR sequences along the channel dimension as the input of the model.
The training iterations were set to 450 epochs with a linear warmup of the first 5 epochs. We trained the model using the Adam optimizer with a batch size of 4 and betas of (0.9, 0.999). The initial learning rate was set to 0.0085 and decayed by multiplied with $(1-\frac{current\_epoch}{max\_epoch})^{0.9}$. We also regularized the training with an $l_2$ weight decay of $1e-5$. 
To reduce the potential overfitting, we further employed several online data augmentations, including random flipping (on all three planes independently), random rotation ($\pm 10^{\circ}$ on all three planes independently), random per-channel intensity shift of [$\pm0.1$] and intensity scaling of [$0.9-1.1$].
We empirically set the base number $K=256$ in all experiments.
We adopted a combination of generalized Dice loss \cite{sudre2017generalised} and binary cross-entropy loss as the loss function.
All experiments were performed based on PyTorch 1.2.0. Since the training of the cascaded HNF-Net requires more than 11 Gb GPU memory, we used 4 NVIDIA Tesla P40 GPUs and 4 NVIDIA Geforce GTX 2080Ti GPUs to train the cascaded HNF-Net and single HNF-Net, respectively.

\begin{table*}[t]
\setlength\tabcolsep{1pt}
\centering
\caption{Segmentation performances of our method on the BraTS 2020 validation set. DSC: dice similarity coefficient, HD95: Hausdorff distance ($95\%$), WT: whole tumor, TC: tumor core, ET: enhancing tumor core. $^*$: ensemble of models trained with 5-fold cross-validation, $^+$: ensemble of models trained with entire training set. }
\scalebox{0.95}{
\begin{tabular}{l|c |c|c|c| c|c| c| c|c|c}
\hline
\multicolumn{1}{l|}{\multirow{2}{*}{Method}}  &\multicolumn{4}{c|}{Dice(\%)}      &\multicolumn{4}{c}{95\%HD(mm)}\\
\cline{2-9}
\multicolumn{1}{c|}{}       &ET             &WT             &TC             &Mean               &ET                 
&WT             &TC             &Mean\\
\hline
Single$^*$                  &0.78492        &0.91261        &0.83532        &0.84428            &26.60476           &4.17888        &5.41503        &12.06622\\
\hline
Single$^+$                  &$\bm{0.78908}$ &0.91218        &0.84887        &0.85004            &$\bm{26.50355}$    &$\bm{4.10500}$ &5.14468        &11.91774\\
\hline
Cascaded$^*$                &0.77647        &0.91084        &0.85631        &0.84787            &26.68954           &4.38397        &$\bm{4.93158}$ &12.00169\\
\hline
Cascaded$^+$                &0.77338        &0.91022        &$\bm{0.85701}$ &0.84687            &29.71248           &4.30247        &4.93369        &12.98288\\
\hline
H$^2$NF-Net             &0.78751        &$\bm{0.91290}$ &0.85461        &$\bm{0.85167}$     &26.57525           &4.18426        &4.97162        &$\bm{11.91038}$\\
\hline
\end{tabular}}
\label{table1}
\end{table*}

\begin{table*}[t]
\setlength\tabcolsep{1pt}
\centering
\caption{The parameter numbers and FLOPs of both single and cascaded HNF-Net.}
\begin{tabular}{l|c |c}
\hline
Method              &Params (M)         &FLOPs(G)\\
\cline{1-3}
Single HNF-Net      &16.85              &436.59\\
\hline
Cascaded HNF-Net    &26.07              &621.09\\
\hline
\end{tabular}
\label{table2}
\end{table*}

\begin{table*}[t]
\setlength\tabcolsep{1pt}
\centering
\caption{Segmentation performances of our method on the BraTS 2020 test set. DSC: dice similarity coefficient, HD95: Hausdorff distance ($95\%$), WT: whole tumor, TC: tumor core, ET: enhancing tumor core.}
\begin{tabular}{l|c |c| c|c| c| c}
\hline
\multicolumn{1}{l|}{\multirow{2}{*}{Method}}  &\multicolumn{3}{c|}{Dice(\%)}      &\multicolumn{3}{c}{95\%HD(mm)}\\
\cline{2-7}
\multicolumn{1}{c|}{}       &ET         &WT         &TC         &ET         &WT         &TC\\
\hline
H$^2$NF-Net                 &0.82775	&0.88790	&0.85375      &13.04490	  &4.53440	  &16.92065\\
\hline
\end{tabular}
\label{table3}
\end{table*}

\noindent\textbf{Inference.}
In the inference phase, we first cropped the original image with a size of $224\times160\times155$, which was determined based on the statistical analysis across the whole dataset to cover the whole brain area but with minimal redundant background voxels.
Then, we segmented the cropped image with sliding patches instead of predicting the whole image at once, where the input patch size and sliding stride were set to $128\times128\times128$ and $32\times32\times27$, respectively.
For each inference patch, we adopted test time augmentation (TTA) to further improve the segmentation performance, including 7 different flipping ({($x$), ($y$), ($z$), ($x$, $y$), ($x$, $z$), ($y$, $z$), ($x$, $y$, $z$)}, where $x$, $y$, $z$ denotes three axes, respectively).
Then, we averaged the predictions of the augmented and partly overlapped patches to generate the whole image segmentation result. 
At last, suggested by the previous work \cite{isensee2018no,jia2020learning}, we performed a post-processing by replacing enhancing tumor with NCR/NET when the volume of predicted enhancing tumor is less than the threshold. 
Based on the results on the validation set, the threshold value was set to 300 and 500 for the single and cascaded models, respectively.

\begin{figure}[tb]
\centering
\includegraphics[width=0.8\textwidth]{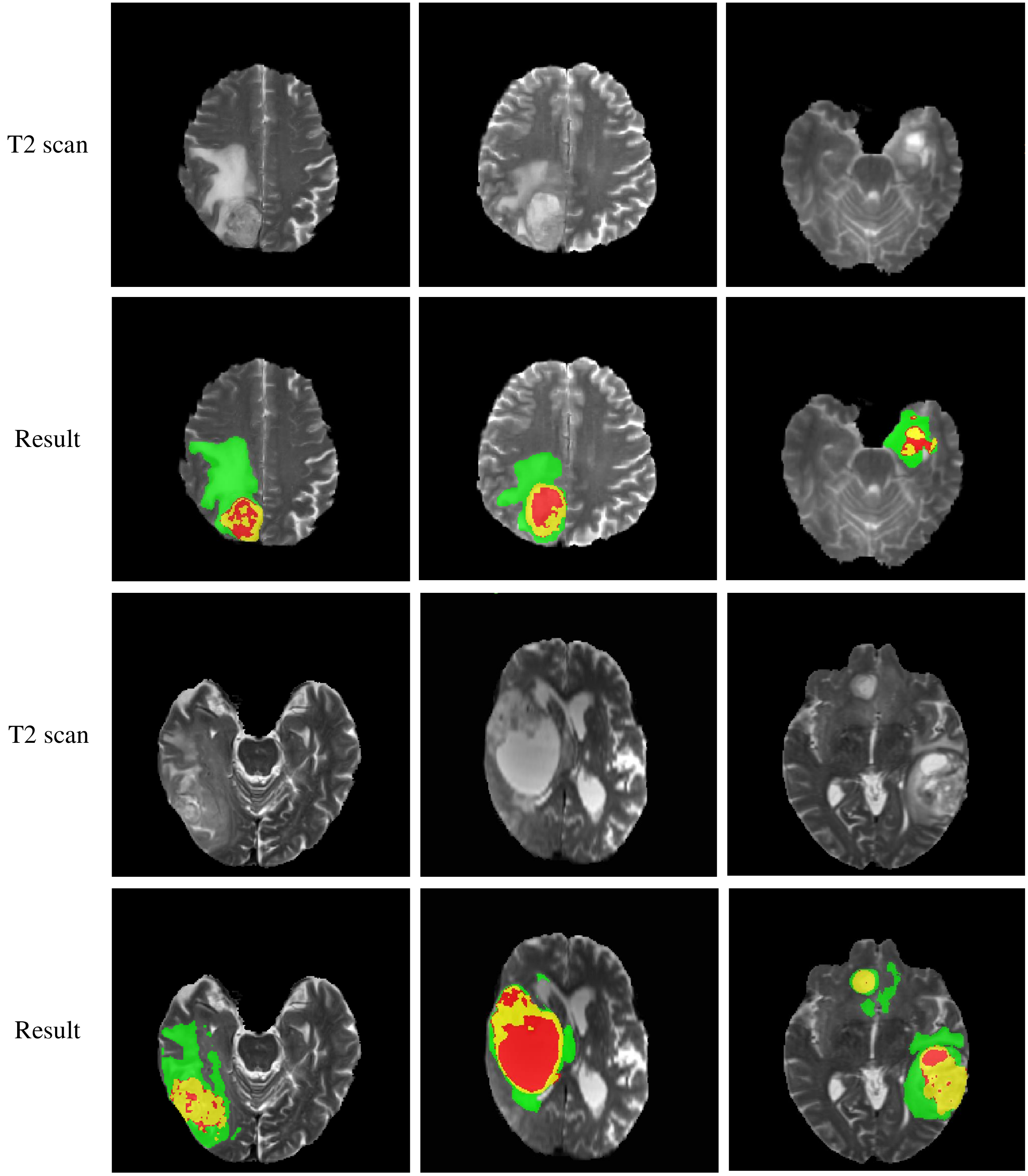}
\caption{Some visual results of our method on the test set. The NCR/NET, ED, and ET regions are highlighted in red, green, and yellow, respectively. } 
\label{fig:fig4}
\end{figure}

\subsection{Results on the BraTS 2020 Challenge Dataset} 
We first evaluated the performance of our method on the validation set, with the results shown in Table \ref{table1}.
All segmentation results were evaluated by the Dice score and 95\% Hausdorff distance (\%95HD) and directly obtained from the BraTS 2020 challenge leaderboard. 
Here Single$^*$ and Cascaded$^*$ denote the ensemble of 5 single models and 5 cascaded models trained with 5-fold cross-validation, respectively. Single$^+$ and Cascaded$^+$ denote the ensemble of 7 single models and 10 cascaded models trained with entire training set, respectively. 
We can find that the single models show better segmentation performance on ET and WT, while the cascaded models have better results on TC.
Besides, our H$^2$NF-Net which has an ensemble of all above 27 models, achieved the best performance on validation set, with a mean Dice score of 0.85167 and a mean \%95HD of 11.91038.
We also provide the parameter numbers and FLOPs of both Single and Cascaded HNF-Net, shown in Table \ref{table2}. 

At last, we further used the ensemble H$^2$NF-Net to segment the test set and the results are shown in Table \ref{table3}. We can observe that our approach achieved average Dice scores of 0.78751, 0.91290, and 0.85461, as well as Hausdorff distances ($95\%$) of 26.57525, 4.18426, and 4.97162 for ET, WT, and TC, respectively, which won the second place out of nearly 80 participants in the BraTS 2020 challenge segmentation task.

Fig. \ref{fig:fig4} visualizes the segmentation results on some scans of the test set. It can be observed that our H$^2$NF-Net can generate convincing results, even on some small regions. 

\section{Conclusion}
In this paper, we propose a H$^2$NF-Net for brain tumor segmentation using multimodal MR imaging, which simultaneously utilizes the single and cascaded HNF-Nets to achieve accurate and robust segmentation. We evaluated our method on the BraTS 2020 validation sets and obtained competitive results. In addition, our H$^2$NF-Net won the second place in the BraTS 2020 challenge segmentation task among near 80 participants, which further demonstrates its superiority and effectiveness.\\

\noindent\textbf{Acknowledgement.}
Haozhe Jia and Yong Xia were partially supported by the Science and Technology Innovation Committee of Shenzhen Municipality, China under Grant JCYJ20180306171334997, the National Natural Science Foundation of China under Grant 61771397, and the Innovation Foundation for Doctor Dissertation of Northwestern Polytechnical University under Grant CX202042.\\

\bibliographystyle{splncs03}
\bibliography{manuscript}
\end{document}